\documentclass[aps,pra,twocolumn,showpacs]{revtex4}

\usepackage{bm}
\usepackage{epsfig}
\usepackage{psfig}

\begin{document}

\title{Bubble domains in disc-shaped ferromagnetic particles}
\author{S. Komineas, C.A.F. Vaz, J.A.C. Bland}
\affiliation{Cavendish Laboratory,
Madingley Road, Cambridge CB3 0HE, United Kingdom.}
\author{N. Papanicolaou}
\affiliation{University of Crete, and Research Centre of Crete,
Heraklion, Crete, Greece.}

\date{\today}

\begin{abstract}
We study the fundamental magnetic states of disc-shaped ferromagnetic particles
with a uniaxial anisotropy along the symmetry axis.
Besides the monodomain, a bidomain state is also identified
and studied both numerically and theoretically.
This bidomain state
consists of two coaxial oppositely magnetized cylindrically   
symmetric domains and remains stable even at zero bias field,
unlike magnetic bubbles in ferromagnetic films.
For a given disc thickness we find the critical radius
above which the magnetization configuration falls into
the bidomain bubble state. The critical radius depends strongly on
the film thickness especially for ultrathin films.
The effect of an external field is also studied and
the bidomain state is found to remain stable over a range
of field strengths.
\end{abstract}

\maketitle

Current developments in experimental techniques allow the engineering
of magnetic particles to very small length scales, typically tens
or a few hundred of nanometres \cite{mesoscopic}.
In such systems the characteristic length scales of the magnetic
domains may be comparable to the system size.
The particles should thus support only a small number of magnetic domains
and these should be usually biased by the symmetry of the particle 
to give a small number of distinct magnetic configurations.
Examples are vortex states observed in disc particles
\cite{mesoscopic,shinjo,wachowiak}
as well as vortex and onion states observed in magnetic rings
\cite{klaui,ross}.
This situation is in marked contrast to a bulk ferromagnet where we
have a large number of magnetic domains with significantly different sizes.

Recently, much research effort has focused on the investigation of
narrow ferromagnetic rings where it is found that single,
highly symmetric domain structures mediate the switching between the
high moment (onion) state and low moment (vortex) states \cite{rothman}.
Motivated by the work on rings, in the present case we search
for the existence of high symmetry stable states in circular disc
structures with perpendicular anisotropy, in view of
the recent development of materials with
a large uniaxial anisotropy \cite{shima,christodoulides} and
experiments in permalloy discs \cite{eames}.
In particular, we identify bidomain states which exist even in the
absence of a bias field but are otherwise the analogues
of the magnetic bubbles observed in ferromagnetic films.
In addition, the bidomain states remain stable for a
range of applied fields and can be manipulated in a controlled manner.
Furthermore, we estimate
the size of the magnetic domains supported
in the particle.
The importance of the present findings is that, again, very simple,
high symmetry domain structures are found to be stable and to mediate
the magnetic switching process, in contrast to the complex behaviour
which usually occurs in small elements.

Static as well as dynamical properties of the magnetization
$\bm{m}$ are governed by the Landau-Lifshitz equation.
The constant length of the magnetization is
normalized to unity: $\bm{m}^2=1$.
An important length scale of the system is the exchange length
\begin{equation}  \label{units}
\ell_{ex} = \sqrt{\frac{A}{2\pi M_0^2}},
\end{equation}
where $A$ is the exchange constant and $M_0$ is the
saturation magnetization. In the following, we shall use $\ell_{\rm ex}$ as
the unit of length.
Another important quantity is the dimensionless quality factor
\begin{equation}  \label{kappa}
 \kappa = \frac{K}{2 \pi M_0^2},
\end{equation}
where $K$ is the anisotropy constant.
The significance of the quality factor can be seen in two important
quantities. First, the domain wall width is $\ell_{\rm ex}/\sqrt{\kappa}$.
Second, $\kappa$ controls the relative strength of the magnetostatic
field which has a demagnetizing effect, with respect to the anisotropy
field which favours alignment along a direction perpendicular
to the film.
We shall suppose here that $\kappa > 1$ which means that the
anisotropy is, in general, stronger than the demagnetising field.

The Landau-Lifshitz equation is the basis for all
our calculations and we are interested only in its static solutions.
We find such magnetic configurations
by a relaxation algorithm the details of which were explained in
\cite{komineas,komineasphd}.
We introduce a Gilbert damping term in the equation
and feed the algorithm with an initial guess state. This
eventually converges to a static solution at a local minimum of the energy
functional.

The most demanding part of the method is by far the calculation of the
magnetostatic field which requires the solution of a boundary
value Poisson problem. This makes the numerical simulation of most realistic
problems practically impossible to achieve in three dimensions.
However, a huge reduction of the numerical calculations is obtained
if we confine
our interest to axially symmetric configurations, and in the rest of this
paper we shall be concerned only with such magnetic states. We expect
that this is not a serious constraint at least for the most basic magnetic
states of small particles which will be our main focus.
Indeed, it is reasonable to assume that the lowest lying states of the system
will have the symmetry of the geometry of the particle.
We call $z$ the axis of symmetry of the disc
which is also the direction of the easy axis.
We go to cylindrical coordinates and suppose that the radial
($m_\rho$), azimuthal ($m_\phi$) and longitudinal ($m_z$) components of the
magnetization vector  are functions of $\rho$ and $z$ only:
$
m_\rho  =  m_\rho(\rho , z),\,
m_\phi  =  m_\phi(\rho , z),\,
m_z  =  m_z(\rho , z).
$

As a first step we calculate the simplest possible state.
This is expected to be a single-domain state
in which all spins are driven by anisotropy and
lie roughly along the symmetry $z$
axis. We use the uniform $\bm{m}=(0,0,1)$ state as an initial guess
in the relaxation algorithm.
This then quickly converges to a quasi-uniform static state, at least for
strong anisotropy $\kappa > 1$.
The magnetization vector deviates from the $z$ axis only around the edges
of the disc.
The quasi-uniform state is a monodomain state and is thus expected to be
the ground state for sufficiently small particles.
The transformation $\bm{m} \to -\bm{m}$ gives a second monodomain state.

As the size of the particle becomes larger it is anticipated that the magnetic
configuration will break up into domains.
We conjecture a bidomain state that is axially symmetric.
This consists of an inner cylindrical domain of ``down'' magnetization
surrounded by the outer domain of ``up'' magnetization.
A domain wall has to separate the two domains.
We shall call these axially symmetric bidomain states
``bubbles'' because they bear some essential similarities
to the so-called magnetic bubbles observed in abundance in
ferromagnetic films.

Ferromagnetic films with a strong perpendicular anisotropy were
studied experimentally and theoretically around the 70s.
These early studies were largely driven by technological interest in
magnetic bubbles whose statics and dynamics were analysed
in detail \cite{slonczewski}.
A magnetic bubble is a circular domain of opposite magnetization
in an otherwise uniformly magnetized film with magnetization
perpendicular to the film.
The presence of an external bias field is essential for the stabilisation
of these structures \cite{thiele,thiele2}.
If the bias field is lifted then the magnetostatic field destroys the
bubble which expands and eventually transforms into stripe domains.
In contrast, the bubble states calculated here for sufficiently
small ferromagnetic particles remain stable even in the absence
of a bias field.

\begin{figure}
  \psfig{file=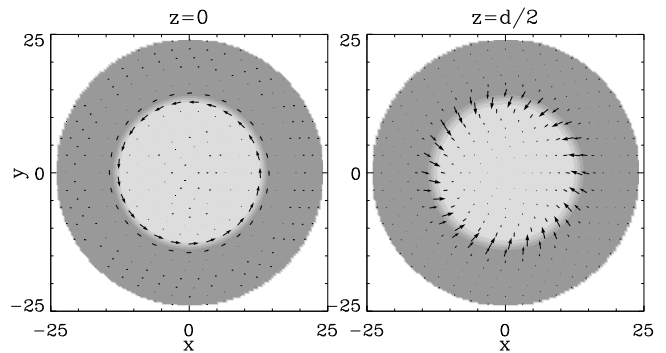,width=8.5cm,bbllx=35bp,bblly=540bp,bburx=540bp,bbury=810bp}
  \caption{
The bubble state illustrated at the middle ($z=0$) plane of the disc and at
the top ($z=d/2$).
The magnetization is such that $m_z=-1$ at the center and $m_z \approx 1$
in the outer domain. The arrows show the projection
of the magnetization on the ($x,y$) plane which has a significant
value at the domain wall. This is a Bloch-like wall in the middle ($z=0$) plane
and it turns almost N\'eel-like at the top and bottom surfaces ($z=\pm d/2$).
The magnetization satisfies
the parity relations $m_\rho(\rho,z)=-m_\rho(\rho,-z),\,
m_\phi(\rho,z)=m_\phi(\rho,-z),\,m_z(\rho,z)=m_z(\rho,-z)$.
}
\label{fig:bubble_image}
\end{figure}

We thus return to our conjecture and proceed to test it numerically.
As a standard example we choose the following values for the quality factor,
the film thickness, and the radius:
\begin{equation}  \label{nominal}
\kappa=2,\quad d = 8\, \ell_{\rm ex}, \qquad R=24\, \ell_{\rm ex},
\end{equation}
and this choice will be explained later in the text.
In the case of the FePt films of Ref. \cite{shima} where
$
M_0  =  1150\; {\rm emu}/{\rm cm}^3,
A  =  10^{-6}\; {\rm erg/cm},
$
the exchange length is $\ell_{ex} = 3.5\, {\rm nm}$ and thus
the values of Eq.~(\ref{nominal}) are translated to
$d=28\, {\rm nm}, R=84\, {\rm nm}$.
The value $\kappa=2$ for the quality factor corresponds to
$K=1.6\times 10^7\; {\rm erg/cm}^3$ for the anisotropy constant,
as is typical in FePt films.
We feed our numerical algorithm with an initial condition which has
the gross features of the bubble state described above
with a domain wall of the Bloch type smoothly connecting the domains.
The algorithm converges to a static bubble state which has a
complicated domain wall structure shown in
Fig.~\ref{fig:bubble_image}.
Also, the magnetization deviates to some extent from the $z$ direction
at the side surface of the particle.
The profile of this structure is sufficiently interesting and deserves
attention. The magnetostatic energy is the driving force here and it
clearly favours a bidomain state with opposite magnetization where
the total magnetization would roughly vanish.
On the other hand, the anisotropy and exchange energies are significant
at the domain wall and they put a tension on it to shrink.
In the final result the bubble has an inner domain with volume
smaller than the outer domain. Thus the total magnetization points
along the symmetry axis and it is nonzero.
As mentioned already, contrary to the situation in films,
we suppose here that no external bias field is present.

The domain wall resembles those discussed
in the literature in related calculations \cite{blake,komineas}.
It is Bloch in the central plane (set at $z=0$ here)
and it progressively becomes N\'eel towards the surfaces.
The N\'eel wall is significantly wider than the Bloch wall.
The radius of the bubble is larger at the centre than near the surfaces, but
this is a small effect.
It is easy to understand that for this type of wall
the magnetostatic energy and the total energy density 
are larger near the surfaces than at the disc centre.
In short, the surfaces disfavour the domain wall and subsequently
also the bubble state.
The final and important result is that, for the parameters (\ref{nominal}),
the bubble state has a lower energy than the monodomain state.

\begin{figure}
  \psfig{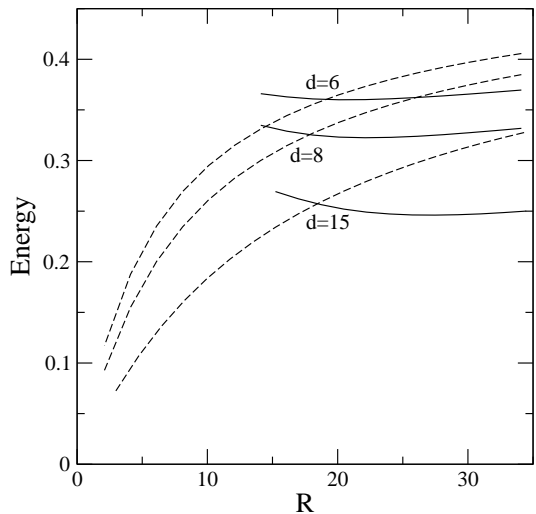}
  \caption{
Energy per unit volume (in units of $2\pi M_0^2$)
of the monodomain (dashed lines) and of the bubble state (solid lines)
as a function
of the disc radius $R$ for three values of the disc thickness $d$
($R$ and $d$ in exchange length units).
The bubble exists only for $R$ greater than a critical radius $R_1$
and it has a lower energy for $R > R_c$ where $R_c$ is yet another
critical radius which corresponds to the intersection of the two
lines for each value of $d$. Both $R_1$ and $R_c$ depend on $d$.
}
\label{fig:EvR}
\end{figure}

We now proceed to a systematic numerical study of the bubble state.
We first fix the disc thickness at $d=8.0$ and vary the radius $R$.
We find a bubble state when the radius is larger than some critical
radius $R_1 \approx 14$.
For smaller radii $R<R_1$ our algorithm always converges to the quasi-uniform
state irrespective of the initial condition.
Our results thus indicate that
the inner bubble domain cannot be sustained if it is too small.
For a disc radius slightly larger than $R_1$ the bubble radius
is small and the total
magnetization of the structure is large.
As the radius of the disc increases the inner bubble domain expands
and the absolute value of the total magnetization decreases.

The energy per unit volume of the bubble state as a function of the disc
radius is given in Fig.~\ref{fig:EvR} for three values of the thickness $d$,
along with the corresponding energy for the monodomain state.
The latter exists as a local minimum of the energy
for any radius up to the largest that we checked.
The energy per unit volume of the bubble is greater than that of
the monodomain state at the lowest radius $R_1$ where the bubble
first appears.
It decreases for larger radii and becomes lower than
that of the monodomain state above a critical radius $R_c$.
It eventually becomes an increasing function of the radius but apparently
remains lower than the energy of the monodomain state for all $R>R_c$.
We consider $R_c$ as marking the size for the break up of the
magnetization configuration into domains.
We could find the bubble as a local minimum of the energy for all radii
$R>R_1$ that we have checked. However, it is expected that a multi-domain
state will eventually set in for a sufficiently large radius,
with energy lower than the energy of both the monodomain and the bubble.

\begin{figure}
 \psfig{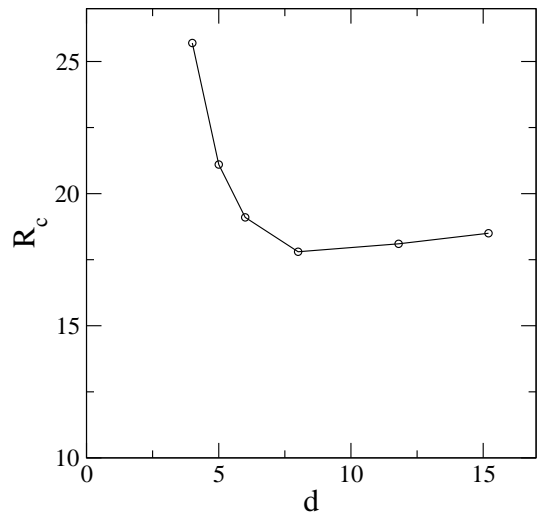}
  \caption{
The critical radius $R_c$ as a function of the thickness of the disc $d$.
For $R > R_c$ the bubble has a lower energy than the monodomain state.
}
\label{fig:Rcvt}
\end{figure}

In order to study the dependence of $R_c$ on the thickness $d$ we have
repeated our calculation for a few values of $d$.
From Fig.~\ref{fig:EvR}
one can extract the critical radii for
three values of $d=6,8,15$.
In Fig.~\ref{fig:Rcvt} we give the $R_c$ as a function of $d$
inferred from six values of $d$.
For small $d$ the critical radius 
significantly exceeds the particle thickness
because surface effects become important and disfavour the formation
of a domain wall.
On the other hand, the critical radius $R_c$ levels off for higher
values of $d$.
The thickness for which $R_c$ attains a minimum appears to be close
to $d=8$ for which $R_c \approx 18$. This is actually the reason for
choosing (\ref{nominal}) as our standard parameters along with the
fact that for $d=8$ the energy of the bubble has a minimum at
$R \approx 24$ as is seen in Fig.~\ref{fig:EvR}.

We have also repeated our calculation for the particle sizes employed
in the experiment of Ref. \cite{eames}
and have confirmed the existence of a bubble state.
However, a detailed quantitative comparison cannot be made before
one determines the
strength of the deposition induced anisotropy in the permalloy
used in the experiment.

Once we have established the existence of a bubble we would like to know
how it behaves under an externally applied field.
Apart from the apparent practical implications, this
is an interesting question also because the
field will affect the intricate balance
of energies that is responsible for the stabilization of the bubble.
The field is applied along the symmetry axis
of the disc, i.e., it is of the form
$\bm{h}_{\rm ext} = (0,0,h_{\rm ext})$.

\begin{figure}
 \psfig{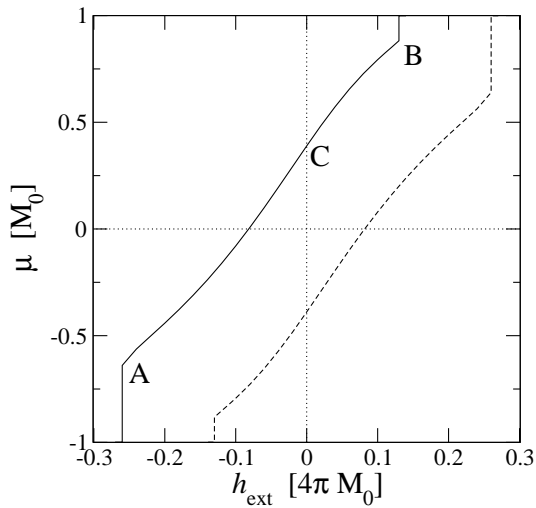}
  \caption{
The total magnetization $\mu$ per unit volume as a function
of the applied field,
for a bubble state in a disc of thickness $d=8$ and radius $R=24$.
}
\label{fig:M-H}
\end{figure}

We apply the field on a particle which is already in a bubble state.
As a specific example, we choose our standard parameter values (\ref{nominal}).
Our results are given in Fig.~\ref{fig:M-H}. 
For $h_{\rm ext}=0$ the total magnetization per unit volume,
$\mu=1/V\int{m_z dV}$, is non-zero.
If we choose the inner domain to point down then $\mu$ is positive
(Point C in Fig.~\ref{fig:M-H}).
Applying a positive external field favours the outer domain, which
expands at the expense of the inner domain. The system does reach
a new equilibrium state which is again a bubble state but with
a smaller radius.
This corresponds to an increased value for $\mu$. For a high enough
magnetic field the bubble becomes too small and it cannot be sustained by the
system. In our example $\mu$ jumps to unity for $h_{\rm ext} > h_B = 0.13$
which signals that the bubble shrinks to zero radius
resulting in a monodomain state with magnetization pointing up (Point B).
On the other hand, if we start from point C but now reduce 
$h_{\rm ext}$ to negative values
the inner domain is favoured and pushes the domain wall
to a larger radius, which
is reflected in a smaller value for $\mu$.
The total magnetization $\mu$ crosses zero for $h_{\rm ext} = -0.08$,
it then becomes negative and eventually
$\mu$ jumps to minus unity for $h_{\rm ext} < h_A = -0.26$
which means that the system is in the monodomain state pointing down (Point A).
Below $h_A$ the domain wall is attracted by the side surface
and is expelled from the disc.
The dashed line in Fig.~\ref{fig:M-H} corresponds to the equivalent
physical situation obtained by the symmetry transformation
$
\bm{m} \to -\bm{m}, \quad \bm{h}_{\rm ext} \to -\bm{h}_{\rm ext}.  \nonumber
$

The bubble is stable in the range $h_A < h_{\rm ext} < h_B$ in which
reversible behaviour occurs. For $h < h_A$ and $h > h_B$
irreversible jumps in the magnetization occur which correspond to the
domain wall being attracted to the edge (point A) or shrink to
the center of the disc (point B).
The size of the jumps in the magnetization
reflect the size of the domain wall being anihilated (larger
when the inner domain expands) and constitutes yet another
example of how the geometry of the element constrains the
shape of the domain wall and the details of the switching
process \cite{vaz,klaui2}.
On the other hand, if a particle with $R > R_c$ is saturated by a strong
{\it in-plane field}, it will eventually relax into a bubble state
after the field is removed \cite{eames}.

In conclusion, we have studied the fundamental states of disc-shaped
magnetic particles with uniaxial anisotropy along the axis of the disc.
A magnetic bubble state has been identified within our
numerical calculation and has been studied in detail.
The bubble is a particularly simple axially symmetric state
and has apparently been observed experimentally \cite{eames}.

\vspace{10pt}

We are grateful to N.R. Cooper, A. Ntatsis, C.A. Ross
and T. Shima for discussions.
This work was supported by EPSRC grant no. GR/R96026/01 (SK).


\end{document}